\documentclass[10pt,letterpaper,twocolumn]{article} 
\usepackage{ol2}
\usepackage{hyperref}
\usepackage{amsmath}

\begin{document}

\twocolumn[ 

\title{A novel method to investigate how the spatial correlation 
of the pump beam affects the purity of polarization entangled states}

\author{Simone. Cialdi$^{1,2,*}$, Davide Brivio$^1$, Andrea
Tabacchini$^{1}$, Ali Mohammed Kadhimi$^{3}$, and Matteo G. A. Paris$^{1,4}$}

\address{
$^1$Dipartimento di Fisica, Universit\`a degli Studi di Milano, I-20133, Milan, Italy \\
$^2$INFN, Sezione di Milano, I-20133 Milan, Italy \\
$^3$Institute of Laser for postgraduate studies, Baghdad University, Iraq \\
$^4$CNISM, Udr Milano, I-20133 Milan, Italy \\
$^*$Corresponding author: simone.cialdi@unimi.it
}

\begin{abstract}
We present an innovative method to address the relation between
the purity of type-I polarization entangled states and the spatial 
properties of the pump laser beam. Our all-optical apparatus is based 
on a spatial light modulator and it offers unprecedented control on the 
spatial phase function of the entangled states. In this way, we demonstrate
quantitatively the relation between the purity of the generated state
and spatial correlation function of the pump beam.
\end{abstract}

 ] 

Spontaneous parametric down-conversion (SPDC) is a crucial process in
the development of quantum technology, and represents one of the most
effective sources of entangled photon pairs and of single photons \cite{kwi95,kwi99}.  
For these reasons the spatial and the spectral properties of the downconverted beams
have been extensively analyzed as a function of the coherence properties 
of the pump beam \cite{gri11, oso07, mot05, lee05}.  
Less attention has been paid to the effect of the spatial properties of 
the pump on the purity of polarization entangled states, especially those 
generated by with type-I parametric down-conversion, since
this may be revealed only by an accurate control of the phase profile of
the output beams.  In this letter, we exploit an all-optical innovative
method based on a spatial light modulator (SLM) to gain an unprecedented control
on the spatial phase function of the generated entangled states and
demonstrate experimentally the relation between the purity of the
generated state and the correlation function of the pump beam.
\par
The downconverted state at the output of the two crystals, 
assuming that the spectra of the pump and of the parametric 
down-conversion are quasi-monochromatic, may be written as:
\begin{align}
 |\Psi\rangle =\frac{1}{\sqrt{2}} \int\int\!\! d \theta_s d \theta_i \,\,
 \hbox{sinc} \left( \frac{1}{2} \Delta k_\parallel L  \right)
 F(\Delta k_{\perp}) \nonumber\\
 \left[ |H, \theta_s \rangle \, |H, \theta_i \rangle \, + e^{i
 \Phi(\theta_s, \theta_i)}|V, \theta_s \rangle \, |V, \theta_i \rangle
 \, \right]\,,
\end{align}
where $L$ is the crystals length and $|P, \theta \rangle$ denote a
single-photon state with polarization $P=H,V$ emitted at angle $\theta$. 
$\Delta k_\parallel$ and $\Delta k_\perp$ are the shifts respect to the 
phase matching condition of the longitudinal and transverse momentum of 
the two photons. The sinc function comes from the integration along the 
longitudinal coordinate inside the crystals, and the function $F$ from 
the integration over the transverse coordinate: denoting by $A_p(x)$ the 
complex amplitude of the pump, we have
$F(\Delta k_{\perp}) = \int\!dx A_p(x)\, e^{i \Delta k_{\perp} x }$.
The phase term $\Phi(\theta_s, \theta_i)$ arises from the optical path 
of the two photons generated in the first crystal inside the second 
crystal, and from the spatial walk-off between the H and the V beams 
of the down-conversion outside the crystals\cite{cia10, kwi09}. 
In general, we may write $\Phi(\theta_s, \theta_i) 
= \phi(\theta_s) +  \phi(\theta_i) +\Phi_{a}$ where, up to first order,
we have $\phi(\theta) \simeq  n^e k L/\cos[(\theta_0 + \theta)/n^e] +
k L \tan[(\theta_0 + \theta)/n^e] \sin(\theta_0 + \theta)
\simeq \frac12 {\phi_0}  + \alpha_0 \theta$, where $n^e$ is the extraordinary 
index of refraction in the second crystal, $k=2\pi/\lambda$, $\theta_0$ 
is the central angle, and $L$ is the crystal length.
The term $\Phi_{a}$ represents the 
additional phase possibly added by any external optical component, e.g.
the SLM. The shifts $\Delta k_{\parallel}$ and $\Delta k_{\perp}$ 
are given by 
\begin{align}
\Delta k_{\parallel} &= k_p - k_s \cos[(\theta_0 + \theta_s)/n^o] - k_i
\cos((\theta_0 + \theta_i)/n^o)\notag \\ 
&\simeq k \theta_0 (\theta_s + \theta_i) = k \theta_0 \theta_+ \nonumber \\
\Delta k_{\perp} &= k_s \sin[(\theta_0 + \theta_s)/n^o] - k_i \sin[(\theta_0 + \theta_i)/n^o] 
\notag \\ & \simeq k (\theta_s - \theta_i) = k \theta_- \,,
\end{align}
where $\theta_+ = \theta_s + \theta_i$ and $\theta_- = \theta_s - 
\theta_i$, and $n^o$ is the ordinary index of refraction. 
Using the new variables the overall phase function rewrites as 
$$\Phi(\theta_-, \theta_+) = \phi_0 + \alpha_0 \theta_+ + \Phi_a\,.$$ 
The purity of 
the state, which in this case equals the visibility, may be written as
\begin{align}
p=\int\!\!\!\int\! \!d \theta_+  d \theta_- \left|\hbox{sinc} 
\left( \gamma \theta_+ \right) 
\,\right|^2 \left| F(k \theta_-) \right|^2 \cos\Phi(\theta_+,
\theta_-)\,,\notag
\end{align}
where $\gamma = \frac{1}{2} k \theta_0 L$. The normalization condition 
is given by $\int\!\!\int\! d \theta_+  d \theta_-
\left|\hbox{sinc} \left( \gamma \theta_+   \right) \right|^2 \left| F(k \theta_-) 
\right|^2 = 1$.
Two cases are of special interest: if $\Phi_{a}=-\phi_0$ the purity
does not depend on $F$ and we obtain the case that is usually described
in the literature \cite{cia10, kwi09}. On the other hand, upon imposing 
$\Phi_{a}=-\phi_0 - \alpha_0 \theta_+  + \beta \theta_-$ one obtains 
$\Phi=\beta \theta_-$, i.e. the purity is now a function of $F$. In this 
second case, using the Wiener-Khinchin theorem, we have:
\begin{align}
p=\!\int\!\! d \theta_-  \left| F(k \theta_-) \right|^2 \cos\beta \theta_- 
\propto \left\langle A^* \left(x+\frac{\beta}{k} \right) A(x)
\right\rangle_x\,, \notag \end{align}
i.e. the purity of the state is proportional to the spatial 
correlation function of the pump beam.
\begin{figure}[h!]
\centerline{\includegraphics[width=0.95\columnwidth]{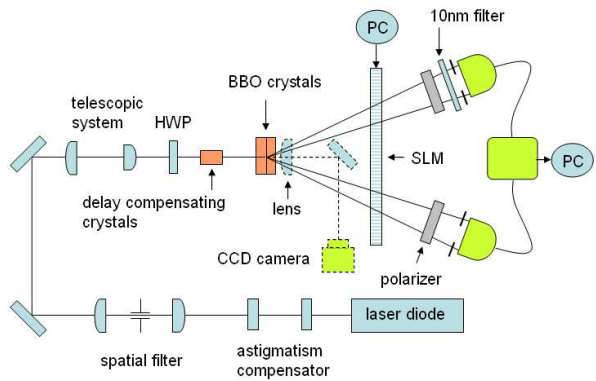}}
\caption{Schematic diagram of the experimental setup. The dashed
part is used to measure the Fourier transform of the pump beam and
it is not present during measurements on the PDC output.
}\label{setup}
\end{figure}
\par
The experimental setup is shown in Fig. \ref{setup}. A linearly polarized cw 
$405$nm diode laser (Newport $LQC405-40P$) passes through two
cylindrical lenses, which compensate beam astigmatism, then a spatial
filter composed by two lens and a pin-hole in the Fourier plane
partially removes the spatial modes, and finally a telescopic system
prepares a beam with the proper beam radius and divergence. A couple of 1-mm
beta-barium borate crystals, cut for type-I down conversion, with
optical axis aligned in perpendicular planes, are used as a source of
polarization and momentum entangled photon pairs with $\theta_0 = 3^o$.
In order to match the above theoretical model we use a compensation crystal
on the pump, which removes the delay time between the vertical and horizontal
polarization \cite{kwi09, cia08}, and put a $10$nm interference filter on the signal
path, in order to reduce the spectral width of the generated radiation.
A spatial light modulator (SLM), which is a liquid crystal phase mask 
($64 \times 10$mm) divided in $640$ horizontal pixels, each wide 
$d = 100 \mu m$, is set before the detectors in order to introduce 
the spatial phase function at $310$mm from the generating crystals \cite{cia10}. 
We also place a window of $5$mm in front of the couplers of the detectors.
A cylindrical lens is placed immediately after the two generating
crystals, whereas a camera sets a the focal distance ($1$m) to obtain the 
square modulus of the Fourier transform of the pump.
\par
In order to complete the theoretical model we have to take into account 
the fact that the spatial coupling is not flat, but rather has a Gaussian profile 
with a FWHM of about $5$mm.  We thus insert this function when tracing 
out the spatial degrees of freedom in order to obtain the polarization 
state and its purity. In addition, there are some elements that introduce 
decoherence not compensable with the SLM: these are the gaps between the 
pixels of the SLM ($3 \mu m$), the imperfect compensation of the delay time, the 
spectral effects of the parametric down-conversion, and the imperfect 
superposition between the amplitudes generated by the two crystals. In 
order to include these effects in the model we assume that the output state is 
described by the mixed state $\rho_ {tot}= m \rho + (1-m) \rho_{mix}$
where $\rho$ is the ideal superposition $\rho = |\Psi \rangle \langle
\Psi |$ and $\rho_{mix}$ the corresponding mixture. The overall purity of 
the state is thus given by $p_{tot}=mp$ where $m$ depends only slightly
on the characteristic of the pump.
\begin{figure}[h!]
\centerline{\includegraphics[width=0.95\columnwidth]{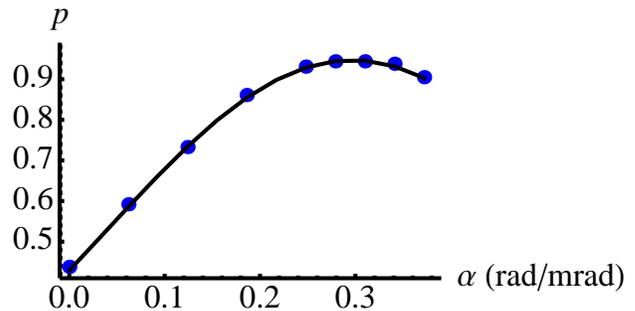}}
\caption{Visibility as a function of $\alpha$ for a beam with 
spot of $220 \mu m$. The phase function imposed by the SLM is
given by $\Phi_a = - \phi_0 -\alpha \theta_+$. Error bars on the
experimental values are within the points. The solid black line 
is the theoretical prediction.}\label{alpha}
\end{figure}
\par
In Fig.~\ref{alpha} we report the typical behaviour of the visibility for 
$\beta=0$, measured as a function of the parameter $\alpha$, which
itself governs the phase function  $\Phi_a = - \phi_0 -\alpha \theta_+$
imposed by the SLM. In this configuration the overall phase function is given 
by $\Phi = (\alpha_0-\alpha) \theta_+$ and thus it is possible to tune
$\Phi_a$ and find 
the optimal value $\alpha=\alpha_0$, which removes the initial phase 
function and maximizes the purity. We find that this value is in good 
agreement with the expected theoretical value.  
The beam has a spot of $220 \mu$m: looking at the Fourier transform we see 
that we are not exactly dealing with a single mode Gaussian profile. However, this 
is not a problem since to fit data we use the square modulus of the Fourier 
transform obtained with the method of the cylindrical lens.
\par
\begin{figure}[htb]
\centerline{\includegraphics[width=1\columnwidth]{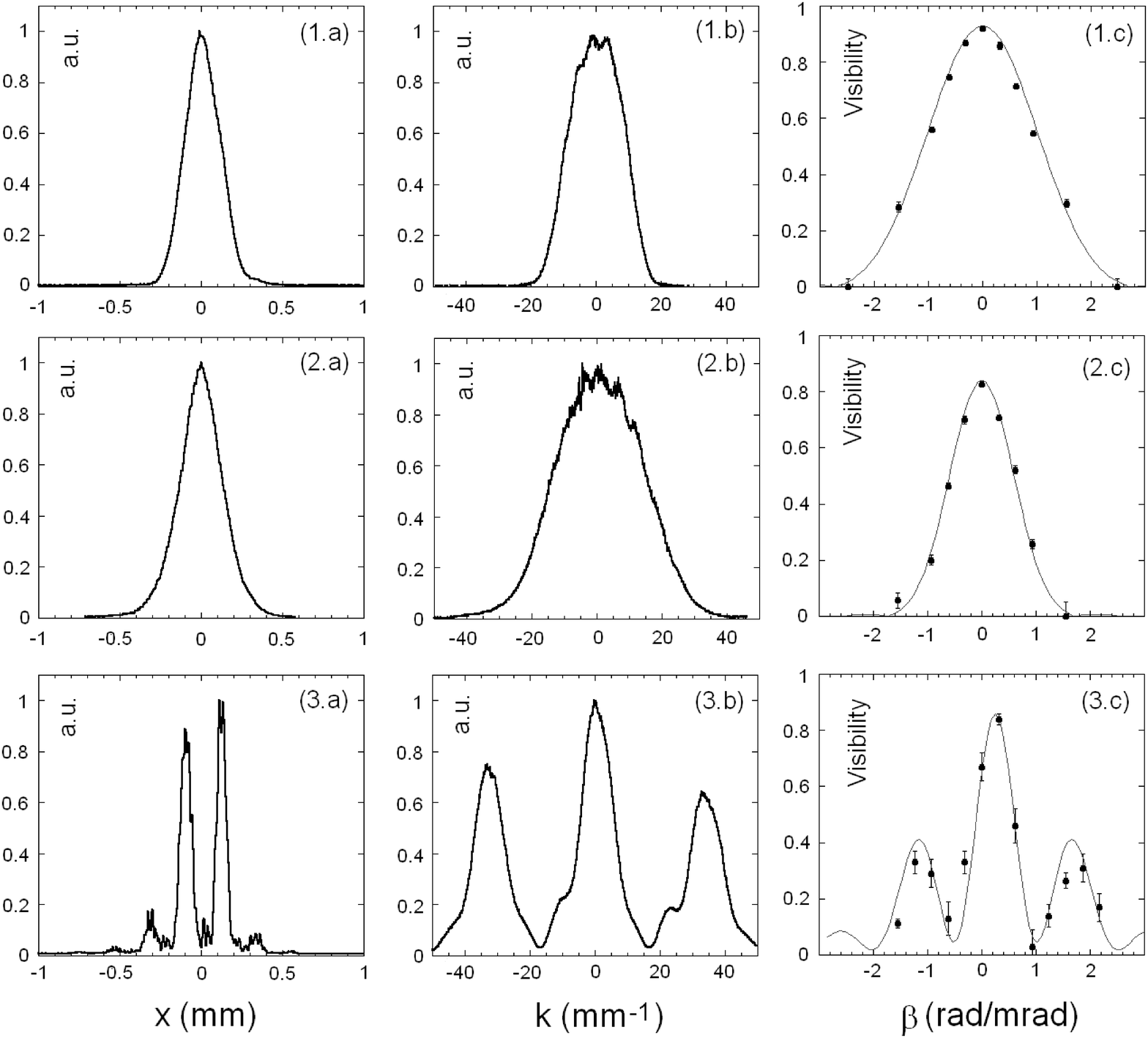}}
\caption{Spatial correlations of the pump beam and purity of the
entangled output in three relevant cases: (first row) collimated pump 
beam of $220 \mu$m, (second row) divergent pump beam of $220 \mu$m, (third row) pump 
beam with two peaks. We report the spatial profile of the pump 
(left column), its Fourier transform (right column), and the visibility
as a function of $\beta$ (right column).
The phase function imposed by the SLM is
given by $\Phi_a = - \phi_0 -\alpha_0 \theta_+ +\beta\theta_-$. Error bars on the
experimental values are within the points. The solid black lines 
are the theoretical predictions (which include the effects of the
walk-off in the two-peaks case).}\label{beta}
\end{figure}

After having found the optimal $\alpha\simeq\alpha_0$ to maximize the 
purity, we then exploit the SLM to impose the phase function 
$\Phi_a = - \phi_0 -\alpha_0 \theta_+
+\beta\theta_-$. In Fig.~\ref{beta} we report the behaviour of the 
visibility as a function of the parameter $\beta$ (right column), together 
with the spatial profile of the pump (left column) and its Fourier 
transform (middle column).
We consider three relevant examples: in the first row we report the 
results obtained with a pin-hole in the spatial filter, in order to 
obtain a quasi single mode Gaussian profile with a spot of $220\mu$m 
on the crystal plane (the same configuration of Fig. \ref{alpha}). 
The visibility has a Gaussian shape, in excellent agreement with the 
theoretical model. In the second case the spatial width is similar 
to the previous case but, as it is apparent from the Fourier transform, 
the beam is divergent (by few mrad). Now the visibility is a Gaussian 
function with a smaller width. In the last example, we place a grid 
with a step of $100 \mu m$ in front of the two generating crystals. In 
the spatial profile we obtain two peaks, and this corresponds to a revival 
in the visibility after a collapse. Since the pump degrees of freedom 
represent a noisy environment for the polarization ones, upon modifying 
the spatial pump profile we are in fact inducing a non Markovian 
dynamics in the polarization degrees of freedom \cite{cia11, bre09}.
The small shift in the bottom right picture (about $0.2$ rad/mrad)
is probably due to an imperfect compensation (of about $30\mu$m) of
the spatial walk-off between the H and V polarization of the pump beam.
\par
In conclusion, we have demonstrated the quantitative relation between the 
purity of type-I polarization entangled states and the spatial properties 
of the pump. In order to obtain this result we exploited the unprecedented 
control of the spatial phase function of the generate states that is achievable 
by the use of a spatial light modulator. Our method may be used for 
entanglement engineering \cite{clu} and purification \cite{cia10}, and it paves the way for 
investigating fundamental effects in non Markovian open systems
\cite{tra}.
\par\noindent

\end{document}